


\def\={\!=\!}
\def\-{\!-\!}
\def\a{\alpha}

\def\d{\partial}
\def\dd{\tilde\partial}
\def\da{^{\dagger}}
\def\e{\eqno}

\def\ie{{\it i.e.}}
\def\pg{paragrassmann}
\def\q{\quad}
\def\qq{\qquad}
\def\qm{quantum mechanics}

\def\t{\theta}
\def\tt{\tilde\theta}

\def\1{{\textstyle{1\over2}}}

\magnification=1200
\vsize=22truecm
\hsize=15truecm
\voffset=0.5truecm
\hoffset=1truecm
\baselineskip=18truept
\lineskip=1pt
\lineskiplimit=0pt
\parskip=6truept


\font\titre=cmbx10 scaled\magstep2
\font\bigletter=cmr10 scaled\magstep2


\baselineskip=12pt

\line{\hfill McGill/93-03}
\line{\hfill hep-th/9305129}
\line{\hfill April 1993}

\vskip 1in
\centerline {\titre Extended}
\vskip 0.05in
\centerline {\titre Fractional Supersymmetric}
\vskip 0.05in
\centerline {\titre Quantum Mechanics}
\vskip 0.75in
\centerline{{{\bigletter S}T\'EPHANE {\bigletter D}URAND}\footnote{$^{*}$}
{E-mail address: durand@hep.physics.mcgill.ca}}
\vskip 0.2in
\centerline{\it Department of Physics}
\centerline{\it McGill University}
\centerline{\it 3600 University Street}
\centerline{\it Montr\'eal, PQ, H3A 2T8, Canada}
\vskip 0.45in

\centerline{To appear in {\it Mod. Phys. Lett.} {\bf A}}
\vskip 0.45in

\centerline{\bf Abstract}
\vskip 0.1in

\noindent
Recently, we presented a new class of quantum-mechanical Hamiltonians
which can be written as the
$F^{\,\rm th}$ power of a conserved charge: $H=Q^F$ with $F=2,3,...\,.$ This
construction, called fractional supersymmetric \qm, was realized
in terms of a paragrassmann variable $\t$ of order $F$, which satisfies
$\t^F=0$.
Here, we present an alternative realization of such an algebra in which
the internal space of the Hamiltonians is described by a tensor product
of {\it two}\/ paragrassmann variables of orders
$F$ and $F-1$ respectively. In particular,
we find $q$-deformed relations (where $q$ are roots of unity) between
different conserved charges.

\vfill
\eject

\baselineskip=18truept

\noindent {\bf 1. Introduction}
\vskip 0.2cm

We have recently presented a new generalization of SUSY \qm\ which we call
{\it fractional}\/ supersymmetric (FSUSY) \qm.$^{[1]}$
In such a construction, the Hamiltonian
is expressed as the
$F^{\,\rm th}$ power of a conserved {\it fractional}\/ supercharge: $H=Q^F$,
with $[H,Q]=0$ and $F=2,3,...\, .$ This is a generalization of the ordinary
SUSY
\qm\ ($F=2$) different from the para-supersymmetric one$^{[2]}$. Moreover,
since the Hamiltonian is the generator of time translation, the FSUSY
transformations associated with $Q$ are the $F^{\,\rm th}$ roots of time
translations. These FSUSY transformations were also described in Ref.~[1] and
an invariant action was provided (a fractional superspace formulation is given
in Ref.~[3]).

In this letter, we present an alternative realization of the algebra defining
the FSUSY \qm. In particular, we have two conserved fractional supercharges
$Q_i$ satisfying the following FSUSY algebras $(i,j=1,2)$:
$$Q_i^F=H, \qq [H,Q_i]=0,        \eqno(1a)$$
$$[Q_i,Q_j]_{q^{i-j}}=0  \eqno(1b)$$
with
$$q^F=1 \qq (q^n\not=1 ~~{\rm for}~~ 0<n<F)     \eqno(2)$$
and where we used the definition
$$[A,B]_\omega\equiv AB-\omega BA. \eqno(3)$$
Note the $q$-deformed-like algebra of $(1b)$. Note also that for $F=2$, the
algebras $(1)$ reduces to the well-known superalgebra
$\{Q_i,Q_j\}=2\delta_{ij}H$.

In Ref.~[1], the construction of FSUSY \qm\ of order $F$ (which we shall
now call {\it minimal}\/ as opposed to the {\it extended}\/ version of the
present work) was realized in terms of {\it one}\/ \pg\ variable $\t$ of
order $F$, which is such that $\t^F=0$. In matrix form, the Hamiltonians
were represented by $F\!\times\!F$ matrices. In contrast, in the present
extended version, the internal space of the quantum-mechanical systems is
described by a tensor
product of {\it two}\/ paragrassmann variables of orders
$F$ and $F-1$ respectively (for $F>2$).
In matrix form, the Hamiltonians are realized in terms of
$F(F-1)\!\times\!F(F-1)$ matrices, and they have
an $F(F-1)$-fold degenerancy (above the ground state). We will see
that these Hamiltonians are also {\it supersymmetric}\/, that is,
they can simultaneously be written as the {\it square}\/ of a conserved
supercharge. Here, we restrict ourself to the one
dimensional case but the construction is
directly applicable for some particular two dimensional systems$^{[1]}$.

In the Sect.~2, we introduce the generalized creation and annihilation
operators
(interpolating between ordinary bosonic and fermionic ones) which are used
to described the internal space of the Hamiltonians.
In Sect.~3, we recall the usual construction of ordinary SUSY \qm.
In Sect.~4, we present the extended FSUSY \qm\ of order 3, and its
generalization to arbitrary order in Sect.~5.

\vskip 0.5cm
\noindent {\bf 2. Paragrassmann variables}
\vskip 0.2cm

In this section, we introduce generalized variables which
interpolate between ordinary bosonic and fermionic ones.
They can be interpreted either as generalized coordinates,
or generalized creation and annihilation operators. The
latter interpretation is more relevant in the present
quantum-mechanical context, but the former
point of view is used in Ref.~[3]
when discussing the fractional superspace formulation of
FSUSY transformations.
The notation, however, will be more reminiscent of the generalized
coordinates interpretation.

We introduce a paragrassmann variable $\t$ of order $M$, and
its derivative $\d\equiv \d/\d\t$, which satisfy
$$\t^M=0, \qq \d^M=0, \qq M=1,2,...   \eqno(4a)$$
$$(\t^{M-1}\not=0, \; \d^{M-1}\not=0) \eqno(4b)$$
In order to be able to recover the 3 different limits which we describe
below (fermionic, bosonic and ``null"), we take the
generalized commutation relation between $\t$ and $\d$ to be
$$[\d,\t]_q=\a(1-q)         \eqno(5)$$
where we have used the definition (3), and where $\a$ is a free parameter
and $q\in\cal C$ a {\it primitive}\/
$M^{\,\rm th}$ root of unity:
$$q^M=1 \qq (q^n\not=1 ~~{\rm for}~~ 0<n<M).  \eqno(6)$$
By a {\it primitive}\/ root, we mean a root satifying the
condition in parentheses; for instance, $q\not=\pm 1$ for $M=4$.
[We will see below that the
condition $(6)$ is actually a consequence of $(4)$ and $(5)$.]
First, note that the null limit $M=1$ $(q=1)$, that is $\t=\d=0$, is
well-defined since the r.h.s of (5) is zero for $q=1$ (and for finite $\a$).
In previous works$^{[4]}$ on \pg\ variables, the r.h.s of $(5)$ was chosen
to be 1 instead of $\a(1-q)$, which is inconsistent in the limit $M=1$
(which we will use here).
Second, note that we recover
the ordinary grassmann case $(q=-1)$ for $M=2$, \ie, $\{\d,\t\}=2\a$.
Third, for some choices of $\a$,
we also recover (within factors) the bosonic case $(q=1)$ for $M\to\infty$.
For instance\footnote{$\da$}
{Other choices are for instance $\a=(M-1)^2/M$ or simply $\a=M/4$, which
lead simultaneously to a well-defined bosonic limit and to the correct
factor for the fermions, \ie\  $\{\d,\t\}=1$.},
for $\a=M$ we find that the r.h.s of (5) is finite and non-zero:
$$\lim_{M\to\infty}M(1-q)=-2\pi i.                   \eqno(7)$$
Strictly speaking, this is the bosonic limit for the first root
$q=exp(2\pi i/M)$.
In the context of a fractional superspace formalism$^{[3]}$, we can choose
to work with real $\t$ and $\d$, whereupon the consistency of the relation
(5) under hermitian conjugation implies that $\a$ must be real.
Moreover, a real $\a$ leads to a real r.h.s of $(10)$ and a real factor in
the r.h.s of $(28)$.
In the following sections, we will not be concerned with the $M\to\infty$
limit, so we let $\alpha$ remain unfixed.

The definition (5) implies
$$[\d,\t^n]_{q^n}=\a(1-q^n)\,\t^{n-1}. \eqno(8)$$
Setting $n=M$ in $(8)$ demonstrates that the consistency of the formalism
requires the condition $(6)$. To see the bosonic limit of that formula,
rewrite $\a(1-q^n)$ as
$$\a(1-q^n)=\a(1-q)(1+q+q^2+...+q^{n-1}) \eqno(9)$$
which leads, for $\a=M$, to $(-2\pi i)n$ in the limit $M\to\infty$.
The definition $(5)$ also implies
$$\sum_{i=0}^{M-1}\,\d^{M-1-i}\,\t^{M-1}\,\d^{i}
=M\a^{M-1}, \eqno(10)$$
as well as the same relation with $\t$ and $\d$ interchanged.
Also note that for a given $q$ of order $M$, one has
$$\sum_{i=1}^{M-1}(1-q^i)=\prod_{i=1}^{M-1}(1-q^i)=M. \eqno(11)$$

We shall also need
the operator $B_{(M)}$ defined as
$$B_{(M)}=\sum_{i=0}^{\infty}c_i\t^i\d^i
=c_0+\sum_{i=1}^{M-1}c_i\t^i\d^i \eqno(12a)$$
with
$$c_0=(1-M)/2 \q {\rm and} \q c_i=[\a^i(1-q^i)]^{-1}
\q (i=1,2,...\, ,M-1). \eqno(12b)$$
This operator has the following properties:
$$[B_{(M)},\t]=\t, \qq [B_{(M)},\d]=-\d. \eqno(13)$$
Note that for $\a=M$ and $M\to\infty$, we have
$c_j\to(i/2\pi)\delta_{j,1}$.

For a given order $M$ which is $prime$, we can actually introduce $M-2$
other fractional derivatives. We thus have $M-1$ derivatives, which we
write as $\d_i$ with
$i=1,2,...,M-1$. With this notation, $\d=\d_1$. They satisfy
$$\d_i^M=0, \qq \d_i^{M-1}\not=0, \qq [\d_i,\t]_{q^i}=\a(1-q^i) \eqno(14)$$
and have the properties
$$[\d_i,\d_{M-i}]_{q^{-i}}=0. \eqno(15)$$
The relation (14) implies both (8) with the substitution $q\to q^i$, and (10)
with the substitution $\d\to\d_i$.
Note that in the SUSY limit $(M=2;q=-1)$ we are left with only one derivative
which satisfies $[\d,\d]_{q^{-1}}=\{\d,\d\}=2\,\d^2=0$ as it must.
For a non-prime $M$, the situation is more complicated.
For instance, for $M=4$, setting $i=2$ in (15) implies $\d_2^2=0$, which
contradicts (14). In other words, $q^i$ is not a primitive root for $i=2$.
Therefore, for a non-prime $M$, there are fewer than $M\-1$ derivatives
$\d_i$, but at least two: $\d\equiv\d_1$ and $\hat\d\equiv\d_{M-1}$. (In
the following, we will use only these two derivatives.)

An $M\!\times\!M$ matrix realization of
$\t$ and $\d$ is given by
$$\t={\pmatrix{0&a_1&0&0&0\cr
               0&0&a_2&0&0\cr
               0&0&0&\ddots&0\cr
               0&0&0&0&a_{M-1}\cr
               0&0&0&0&0\cr}}, \qq
  \d={\pmatrix{0&0&0&0&0\cr
               b_1&0&0&0&0\cr
               0&b_2&0&0&0\cr
               0&0&\ddots&0&0\cr
               0&0&0&b_{M-1}&0\cr}} \e(16a)$$
with the constraint (no summation on $i$)
$$a_i b_i =\a(1-q^{-i}).     \eqno(16b)$$
Note that in general $\d\not=\t\da$.
As for ${\hat\d}$, it is given by
$$  \hat\d={\pmatrix{0&0&0&0&0\cr
               \hat b_1&0&0&0&0\cr
               0&\hat b_2&0&0&0\cr
               0&0&\ddots&0&0\cr
               0&0&0&\hat b_{M-1}&0\cr}}
\q {\rm with}\q {\hat b}_i=-q^i\,b_i.  \eqno(17)$$
In this matrix realization, $B_{(M)}$ is found to be the third component
of the
spin-$(M-1)/2$ representation of the rotational group:
$$B_{(M)}=J_3^{[(M-1)/2]}. \eqno(18)$$

\vskip 0.5cm
\noindent {\bf 3. Ordinary SUSY quantum mechanics}
\vskip 0.2cm

Let us first recall the usual construction of
one-dimensional SUSY quantum mechanics.
We introduce the bosonic operators $a$ and $a\da$:
$$a=[p+iW(x)]/\sqrt2, \qq a\da=[p-iW(x)] /\sqrt2         \eqno(19)$$
which satisfy
$$[a\da,a]={d\over dx}W(x)\equiv W'(x) \eqno(20)$$
where $p=-id/dx$. We need an ordinary grassmann variable $\t$
and its derivative $\d$, which are interpreted as fermionic creation and
annihilation operators. For $M=2$, the commutation relations (4-6) give
$$\{\d,\t\}=2\a, \qq \t^2=\d^2=0. \eqno(21)$$
The conserved supercharge $Q$ and Hamiltonian $H$ satisfying
$$Q^2=H, \qq [H,Q]=0, \eqno(22)$$
are given by
$$Q=\d\,a+e\,\t a\da \eqno(23a)$$
$$H={1\over2}\left(p^2+W^2\right)+W'\cdot S \eqno(23b)$$
where
$$S=-\textstyle{1\over2}+e\,\t\d=B_{(2)} \eqno(24)$$
and $e^{-1}=2\a$. With the realization $(16)$, we find
$$S=\textstyle{1\over2}\sigma_3
={1\over2}\pmatrix{1&0\cr0&-1\cr}.\eqno(25)$$
Thus the Hamiltonian $(23b)$ describes a ``spin-1/2" particle
moving in a potential $W^2/2$ and a ``magnetic field" $W'$.
Moreover, with the choice $\a=1/2$ ($e=1$), we may choose $\t=\sigma_{+}$
and $\d=\sigma_{-}$, \ie, $\d=\t\da$.

\vskip 0.5cm
\noindent {\bf 4. Extended FSUSY quantum mechanics of order 3}
\vskip 0.2cm

To describe the internal space of SUSY and {\it minimal}\/ FSUSY$^{[1]}$
Hamiltonians,
only {\it one}\/ type of variable is used: a \pg\ variable of order $F$.
For the {\it extended}\/ FSUSY case, the situation is different since
{\it two}\/ different types of internal-space variables are needed
simultaneously: one paragrassmann variable of order $F$, and another of
order $F-1$; the variables of different order commute with each other.
(The tensor product of these two types of variables may equivalently be
seen as only
{\it one}\/ internal-space variable.)
Thus, for the case $F=3$, we need both a set of ordinary grassmann variables
$(\tt,\dd)$ [which satisfy (21), \ie, (4-6) with $M=2$],
and a set of paragrassmann variables $(\t,\d)$ of order 3 [which satisfy
(4-6) with $M=3$]. Hence we have:
$$\eqalignno{&\tt^2=\dd^2=0, \qq \{\dd,\tt\}=2\a' &(26a)\cr
             &\t^3=\d^3=0, \qq [\d,\t]_q=\a(1-q) &(26b)\cr
             &\q\qq\,\, q^3=1 \q (q\not=1) &(26c)\cr
             &[\t,\tt]=[\t,\dd]=[\d,\tt]=[\d,\dd]=0. &(26d)\cr}$$
[Except on $W(x)$, the prime does not indicate a derivative.]
In particular, from $(26bc)$, we have
$$\d^2\t^2+\d\t^2\d+\t^2\d^2=3\a^2 \eqno(27)$$
and
$$\d^2\t+\d\t\d+\t\d^2=(3\a)\,\d     \e(28)$$
(as well as the same equations with the interchange $\t\leftrightarrow\d$).
The formula (27), which is a particular case of $(10)$, is useful in the
present FSUSY context, whereas the formula (28) is used in
rewriting\footnote{$\da$}
{In the context of PSUSY \qm $^{[2]}$, the \pg\ variables of order 3 are
traditionally taken to satisfy $\d^2\t+\t\d^2=2\d$ and $\d\t\d=2\d$. However,
only the sum of these two relations, \ie\ (28), is really needed.}
the PSUSY quantum mechanics$^{[2]}$ in terms of these paragrassmann
variables$^{[5]}$.

We define the ``square root" of the bosonic operator $a$ of $(19)$ as
$$a^{1/2}\equiv\dd+e'\tt a, \qq (a^{1/2})^2=a, \eqno(29)$$
with $(e')^{-1}=2\a'$. Now we introduce the FSUSY Hamiltonian $H$
of order $3$ and the associated conserved charge $Q$ which satisfy
$$Q^3=H, \qq [H,Q]=0. \eqno(30)$$
They are given by
$$Q=\d\,a^{1/2}+e\,\t^{2}a\da \eqno(31a)$$
$$H={1\over2}(p^2+W^2)+W'\cdot S\eqno(31b)$$
where
$$\eqalignno{S&=-\textstyle{1\over2}+
e e'\,\d\t^2\d\cdot\dd\tt+e\,\t^2\d^2&(32a)\cr
              &={1\over2}[B_{(3)}-B_{(2)}+B_{(3)}^2\cdot B_{(2)}]
                          &(32b)\cr}$$
and with $e^{-1}=3\a^2$.
With the particular realization $(16)$ and understanding the dot in $(32)$
as the tensor product of the 3$\times$3 matrix by the 2$\times$2 matrix,
we get for $S$ the following 6$\times$6 matrix:
$$S={1\over2}{\pmatrix{\sigma_3&0&0\cr0&-\sigma_3&0\cr0&0&\sigma_3\cr}}
                                                            \eqno(33)$$
where $\sigma_3$ is the third Pauli matrix given in $(25)$.
Thus, in this matrix form, the Hamiltonian $(31b)$ is diagonal and is a sum
of 3 ordinary SUSY Hamiltonians, each of them describing a spin-1/2 particle
in a potential $W^2/2$ and interacting with a ``magnetic field" $\pm W'$.
With a similarity transformation, it is easy to change the sign of the middle
$\sigma_3$ of $(33)$ [by interchanging the two middle diagonal elements].
Therefore, the Hamiltonian $(31b)$ is also supersymmetric; that is, it can
also be written as the square of a charge: $H=\tilde Q^2$ with
$\tilde Q=T[(\dd\otimes{\bf 1}_3)a+e'(\tt\otimes{\bf 1}_3)a\da]T^{-1}$ where
${\bf 1}_3$ is the 3$\times$3-identity in the space of the variables of order
3, and $T$ is some similarity transformation.

Now we want to introduce another conserved fractional charge ${\hat Q}$ of
order 3. For that purpose, we need
the second fractional derivative ${\hat\d}$ which satisfies
$${\hat\d}^3=0, \qq [{\hat\d},\t]_{q^{-1}}=\a(1-q^{-1}). \eqno(34)$$
It has the property
$$[{\hat\d},\d]_q=0.  \eqno(35)$$
The relations (34) and (35) are particular cases of (14) and (15).
Introducing ${\hat Q}$ as
$${\hat Q}=(-1)^{1/3}\big[{\hat\d}\,a^{1/2}-e\,\t^2 a\da\big], \eqno(36)$$
we have the following algebra
$${\hat Q}^3=H, \qq [H,{\hat Q}]=0, \eqno(37a)$$
$$[{\hat Q},Q]_q=0. \eqno(37b)$$
The relation
$(37b)$ is valid only in matrix realizations, that is, with $\hat\d$ given
by (17) (in addition to the matrix realizations of $\t$ and $\d$).
With the notation $Q_1\equiv Q$ and $Q_2\equiv {\hat Q}$, we can rewrite
$(37)$ as in $(1)$ with $F=3$. We can also introduce 3 ``ladder" operators
which cube to zero (see next section).

\vskip 0.5cm
\noindent {\bf 5. Extended FSUSY quantum mechanics of arbitrary order}
\vskip 0.2cm

Now we turn to the general case.
To distinguish between \pg\ variables of different order, we introduce the
notation
$\d_{(M)}$, $\t_{(M)}$ and $\a_{(M)}$ for the
variables ($\d,\t$) of order $M$ and the associated $\a$ defined through
$[\d_{(M)},\t_{(M)}]_q=\a_{(M)}(1-q)$ with $q^M=1$.
We shall now need the ``$p^{\rm th}$ root" of the bosonic operator $a$ of
$(19)$:
$$a^{1/p}\equiv\d_{(p)}+e'\,\t_{(p)}^{p-1}a,\qq (a^{1/p})^p=1,\eqno(38)$$
with $(e')^{-1}=p\,\a_{(p)}^{p-1}$.
Note that in order to recover $a^{1/1}=a$, we need the null limit
$M=1$ of (4). This will be useful when looking for the SUSY limit of the
general case.

Now we introduce the FSUSY Hamiltonian $H_{(F)}$
of order $F$ and the associated conserved charge $Q_{(F)}$:
$$\eqalignno{&Q_{(F)}=\d_{(F)}a^{1/(F-1)}+e\,\t_{(F)}^{F-1}a\da,
\qq e^{-1}=F\a_{(F)}^{F-1}   &(39a)\cr
  &H_{(F)}={1\over2}\left(p^2+W^2\right)+W'\cdot S_{(F)} &(39b)\cr}$$
where
$$S_{(F)}=-{1\over2}
+e e'\sum_{i=1}^{F-1}\bigg\{\d_{(F)}^{F-1-i}\,\t_{(F)}^{F-1}\,\d_{(F)}^{i}
\,\sum_{j=0}^{i-1}\,\d_{(F-1)}^{F-2-j}
\,\t_{(F-1)}^{F-2}\,\d_{(F-1)}^{j}\bigg\}.\eqno(40)$$
They satisfy the following fractional supersymmetry algebra:
$$Q_{\scriptstyle{(F)}}^F=H_{(F)}, \qq [H_{(F)},Q_{(F)}]=0. \eqno(41)$$
Note how we recover the SUSY case for $F=2$ using $a^{1/1}=a$ in $(39a)$.
With the realization (16), we get for $S$ an $F(F-1)\!\times\!F(F-1)$
diagonal matrix with an equal number of
$\pm1$ entries. Therefore, the Hamiltonian $(39b)$ is also SUSY
and the same discussion as for the $F=3$ case applies. In particular, the
spectrum of the harmonic oscillator is $F(F-1)$ degenerate except for the
ground state which has half the degeneracy.

Now we drop the $(F)$ subscripts. Using the other fractional derivatives (14),
one may write $F-1$ conserved charges (fewer for a non-prime $F$)
satisfying $(41)$; it suffices to replace $\d$ in $(39a)$ by $\d_i$.
Rather, let us introduce the following charge (valid for both prime and
non-prime $F$):
$${\hat Q}=(-1)^{1/F}\big[{\hat\d}\,a^{1/(F-1)}-e\,\t^{F-1}a\da\big]
\eqno(42)$$
where we recall the notation ${\hat\d}\equiv\d_{F-1}$. We have the following
algebra:
$${\hat Q}^F=H, \qq [H,{\hat Q}]=0, \eqno(43a)$$
$$[{\hat Q},Q]_q=0. \eqno(43b)$$
The $q$-deformed relation $(43b)$ holds only if one imposes the additional
constraint $\t^{F-1}(\d+q\,{\hat\d})=0$, which is automatically fullfilled
in the matrix realization $(16)$ and $(17)$.
For $F=2$, this constraint is trivial since $\hat\d=\d$ and $q=-1$.

Using the notation $Q_1\equiv Q$ and $Q_2\equiv {\hat Q}$, we may combine
the equations $(41)$ and $(43)$ either as in $(1)$, or in the following way
$(r_i=1,2)$:
$$\{Q_{r_1},Q_{r_2},...\,,Q_{r_F}\}=F\,\delta_{r_1 r_2...r_F}\,H \eqno(44)$$
where $\delta_{r_1 r_2...r_F}=\prod_{i=1}^{F-1}\delta_{r_i,r_{i+1}}$
and where we used the following definition of the multilinear
product:
$$\{X_1,X_2,...\,,X_F\}\equiv X_1X_2...X_F
\,+~{\it cyclic~permutations~of~the}~X_i.\eqno(45)$$
[So, the right hand side of $(45)$ contains
$F$ terms.] This product is a generalization of
the anti-commutator. Note that from $(44)$, we recover for $F=2$
the well-known superalgebra
$$\{Q_i,Q_j\}=2\,\delta_{ij}\,H, \qq i,j=1,2. \e(46)$$

We can also introduce $F$ ``ladder" operators. They are given by
$${\bf Q}_i=Q+q^i(-1)^{-1/F}\hat Q, \qq i=1,2,...\,,F  \e(47)$$
and satisfy
$$\{{\bf Q}_{r_1},{\bf Q}_{r_2},...\,,{\bf
Q}_{r_F}\}=F\,(1-q^{r_1+r_2+...+r_F})\,H.    \e(48)$$
In particular, we have ${\bf Q}_i^F=0$ (but ${\bf Q}_i^{F-1}\not=0$). For
$F=2$, we recover the SUSY case:
$${Q}_{\pm}^2=0, \qq \{{Q}_+,{Q}_-\}=4H,   \e(49)$$
with ${Q}_{\pm}=Q\pm i\hat Q$ (\ie, with the notation
${Q}_+\equiv {\bf Q}_2$ and ${Q}_-\equiv {\bf Q}_1$). In the
construction (47-48), we can include the case $F=1$.

\vfill\eject

\centerline{\bf Acknowledgments}
\vskip 0.2cm

This work is supported in part by a fellowship from
the Natural Sciences and
Engineering Research Council (NSERC) of Canada.


\vskip 0.5cm
\centerline{\bf References}
\vskip 0.2cm

\par
\frenchspacing

\item{[1]}
S. Durand, ``Fractional Supersymmetry and Quantum Mechanics",
preprint McGill/92-54, April 1993, hep-th/9305128.

\item{[2]}
V.A. Rubakov and V.P. Spiridonov,
{\it Mod. Phys. Lett.} {\bf A3}, 1337 (1988);
S. Durand, M. Mayrand, V.P. Spiridonov and L. Vinet,
{\it Mod. Phys. Lett.} {\bf A6}, 3163 (1991).

\item{[3]}
S. Durand, ``Fractional Superspace Formulation of Generalized Mechanics",
preprint McGill/93-06, May 1993, hep-th/9305130.

\item{[4]}
C. Ahn, D. Bernard and A. LeClair, {\it Nucl. Phys.} {\bf B346}, 409 (1990);
S. Durand, {\it Mod. Phys. Lett.} {\bf A7}, 2905 (1992);
A.T. Filippov, A.P. Isaev and A.B. Kurdikov,
{\it Mod. Phys. Lett.} {\bf A7}, 2129 (1992);
``On Paragrassmann Differential calculus", hep-th/9210075;
``Paragrassmann Extensions of the Virasoro Algebra", hep-th/9212157.

\item{[5]}
S. Durand, in preparation.

\par

\vfill
\end